\documentclass[aps,pra,twocolumn,superscriptaddress]{revtex4-2}
\usepackage{amsmath,amssymb}
\usepackage{graphicx}
\usepackage{color}
\pdfoutput=1
\usepackage{bm}
\usepackage{bbm}
\usepackage{amsmath, amsfonts}
\usepackage{amssymb}
\usepackage{xcolor}
\usepackage[colorlinks=true,allcolors=blue]{hyperref}
\usepackage{cleveref}
\usepackage{mathtools}
\begin{document}

\title{Readout-induced degradation of transmon lifetimes: {interplay of TLSs and qubit spectral reshaping}
}
\author{Ziwen Huang}
\thanks{These authors contributed equally to this work.}
\address{Amazon Center for Quantum Computing, Pasadena, California 91106, USA}
\author{Jimmy Shih-Chun Hung}
\thanks{These authors contributed equally to this work.}
\address{Amazon Center for Quantum Computing, Pasadena, California 91106, USA}
\author{\mbox{Mouktik Raha}}
\address{Amazon Center for Quantum Computing, Pasadena, California 91106, USA}
\author{\mbox{{Ming-Han Chou}}}
\address{Amazon Center for Quantum Computing, Pasadena, California 91106, USA}
\author{\mbox{Harry Levine}}
\address{Amazon Center for Quantum Computing, Pasadena, California 91106, USA}
\address{Department of Physics, University of California, Berkeley, California 94720, USA}
\author{\mbox{Alex Retzker}}
\address{Amazon Center for Quantum Computing, Pasadena, California 91106, USA}
\address{Racah Institute of Physics, The Hebrew University of Jerusalem, Jerusalem 91904, Givat Ram, Israel}
\author{\mbox{Connor T. Hann}}
\address{Amazon Center for Quantum Computing, Pasadena, California 91106, USA}
\author{David Hover}
\address{Amazon Center for Quantum Computing, Pasadena, California 91106, USA}
\author{\mbox{Fernando G.S.L. Brand\~{a}o}}
\address{Amazon Center for Quantum Computing, Pasadena, California 91106, USA}
\address{Institute for Quantum Information and Matter, California Institute of Technology, Pasadena, California 91125, USA}
\author{\mbox{Aashish A. Clerk}}
\address{Amazon Center for Quantum Computing, Pasadena, California 91106, USA}
\address{Pritzker School of Molecular Engineering, University of Chicago, Chicago, Illinois 60637, USA}
\author{Arbel Haim}
\address{Amazon Center for Quantum Computing, Pasadena, California 91106, USA}
\author{Oskar Painter}
\address{Amazon Center for Quantum Computing, Pasadena, California 91106, USA}
\address{Institute for Quantum Information and Matter, California Institute of Technology, Pasadena, California 91125, USA}
\address{Thomas J. Watson, Sr., Laboratory of Applied Physics, California Institute of Technology, Pasadena, California 91125, USA}

\begin{abstract}
Measurement backaction degrades dispersive readout of superconducting qubits even at modest drive strengths, often via the reduction of qubit lifetimes during readout. In this work, we theoretically and experimentally study this degradation and show how it can result from the interplay between detuned two-level systems (TLSs) and a drive-renormalized qubit spectrum. For modest to strong readout, the qubit emission spectrum becomes non-Lorentzian and depends sensitively on the readout drive frequency (even when measurement rate is fixed). We combine the readout-modified qubit emission spectrum with time-dependent perturbation theory to predict qubit lifetimes in the presence of a TLS bath. Master equation simulations and experimental measurements on a frequency-tunable transmon confirm these predictions quantitatively. In particular, we find that driving at the resonator frequency associated with the qubit ground state yields the narrowest qubit emission spectrum and the least lifetime degradation for a fixed measurement rate, providing a practical guideline for optimizing readout protocols in future quantum processors.
\end{abstract}

\maketitle

\section{Introduction}

\begin{figure}[h!!]
    \centering
    \includegraphics[width=0.87\linewidth]{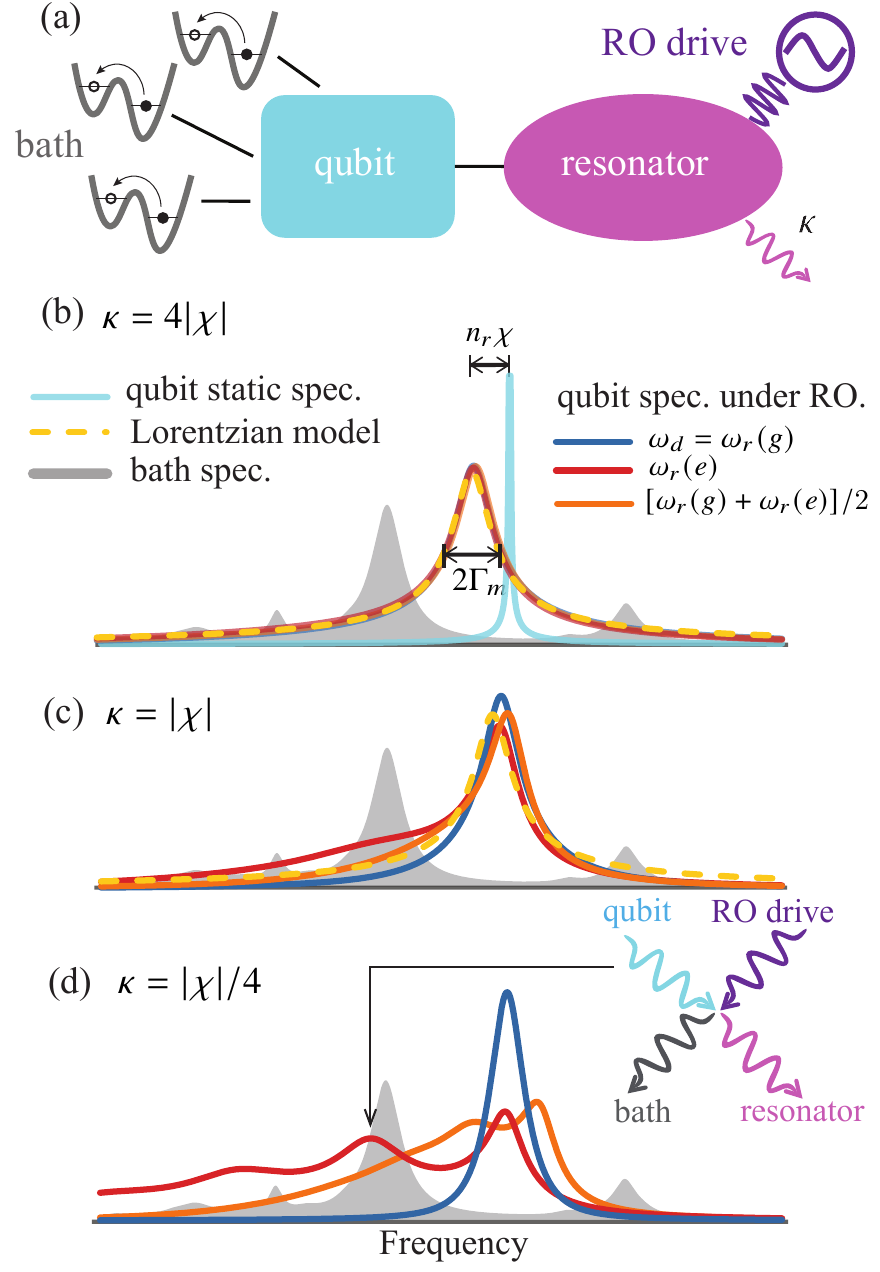}
    \caption{Qubit linewidth broadening in different qubit-resonator coupling regimes. (a) A schematic of the system comprising a superconducting qubit coupled to a readout resonator, which is driven by a readout (RO) pulse and connected to an environmental bath containing TLSs. (b) In the weak-pull limit ($|\chi|\ll \kappa$), the qubit spectrum under readout broadens with width proportional to the photon number and the dispersive shift. This broadening enables interaction with previously detuned TLSs in the bath (gray shading). (c) For a conventional $|\chi|/\kappa$ ratio, the qubit spectrum already deviates from the Lorentzian prediction (dashed line for Lorentzian prediction for center readout frequency). Different colors correspond to three choices of readout frequencies. (d) For a large $|\chi|/\kappa$ ratio, the qubit spectrum exhibits distinctly different lineshapes for each readout frequency choice. For all cases in (c) and (d), the measurement rate is kept constant across the three frequency choices. The inset in (d) depicts the four-wave mixing process associated with the peak marked by the black arrow (see Sec. II.A for a detailed explanation).}
    \label{fig:cartoon}
\end{figure}

Superconducting quantum processors often rely on fast and reliable measurement, typically done with dispersive readout \cite{Blais_cQED,Wallraff_Nature,Blais_CircuitQED_Review}. Modern implementations routinely achieve readout times of a few hundred nanoseconds with errors below 1\% \cite{Krantz_QuantumEngineersGuide, Jeffrey_FastAccurate, Walter_RapidHighFidelity, Heinsoo_MultiplexedReadout,Google_Willow,IBM_utility,Bengtsson_model_based}. Still, faster and more accurate readout is indispensable for building resource-efficient error-corrected quantum computers \cite{Fowler_QEC,ibm_qldpc}. In the conventional model of dispersive readout, faster measurement can be achieved by increasing drive power \cite{Krantz_QuantumEngineersGuide,Blais_cQED}, which should in turn suppress readout error by reducing the time over which the qubit is susceptible to environmental decoherence. In practice, however, many studies observe either a saturation or a reduction of readout fidelity as drive power increases. 

This conventional model fails because additional error channels emerge at higher drive strengths. Recent research has identified two classes of mechanisms responsible for this degradation. The first, known as measurement-induced state transitions (MIST) \cite{Google_MIST_2016,Google_MIST,BlaisIonization,Devoret_DUST,Yale_Hazra_MIST}, is intrinsic to the transmon-resonator system: the readout drive enables spurious, unintended transitions to higher transmon states. The second class involves modes external to the transmon-resonator system (e.g., two-level systems), where the readout drive modifies the qubit's interaction with these external modes, leading to enhanced qubit relaxation or sometimes excitation \cite{IBM_readout_zeno,Murch_zeno,Kou_fluxonium, Devoret_DUST}. The latter can occur at smaller drive amplitudes and, due to the unpredictable nature of TLSs, is often more difficult to mitigate. 

In this work, we present a focused study of the second mechanism, specifically on an important interplay that has been largely overlooked. The prevailing picture for understanding this mechanism is that measurement-induced dephasing broadens the qubit emission spectrum, potentially enabling previously off-resonant TLSs to interact with the qubit and suppress its lifetime \cite{IBM_readout_zeno,Kou_fluxonium,Murch_zeno}. The broadening is generally assumed to be determined by the measurement rate, the rate at which the two qubit states are resolved in the output signal. Examining this picture more carefully, we show that the drive-induced spectral broadening and qubit-TLS interaction depend critically on the choice of readout drive frequency for a fixed measurement rate: choosing certain optimal drive frequencies yields significantly less $T_1$ degradation than others.

In the weak-pull limit, where the dispersive shift $\chi$ is much smaller than the resonator loss rate $\kappa$, this spectral renormalization reduces to a simple ac Stark shift and Lorentzian broadening of the qubit linewidth, as shown in Ref.~\cite{IBM_readout_zeno} [see illustration by the dashed line in Fig.~\ref{fig:cartoon}(b)]. In this limit, knowledge of measurement rate and resonator average photon number is sufficient for determining the readout-modified qubit spectrum. However, many modern superconducting qubit architectures operate at larger $|\chi|/\kappa$ ratios \cite{cqc_dual_rail_exp, cqc_dual_rail_theor,Sank_Ross,IBM_utility,Bengtsson_model_based}, where the simple Lorentzian broadening model is insufficient and a more general theory of the qubit's spectral renormalization under readout is needed. To treat this more general case, we used the theory developed in Refs.~\cite{Gambetta_num_split_theor, Clerk_num_split}, commonly referred to as number splitting theory (a name that highlights the most extreme scenario, where the qubit spectrum splits into discrete peaks [see the red curve in Fig.~\ref{fig:cartoon}(d)]). Two important features of the general qubit spectrum are absent from the Lorentzian model: the spectral shape is non-Lorentzian even for modest $|\chi|/\kappa$ ratios [see comparison in Fig.~\ref{fig:cartoon}(c)], and it depends sensitively on the readout drive frequency $\omega_d$ with the measurement rate held fixed [see for example Fig.~\ref{fig:cartoon}(d)]. Since the qubit decay rate depends on the overlap between the qubit emission spectrum and the TLS noise spectrum, different choices of $\omega_d$ lead to different magnitudes of $T_1$ degradation. This dependence on readout drive frequency has not been previously explored or experimentally tested, yet it is critical for devices operating beyond the weak-pull limit.

By incorporating these features, we develop a generalized model for readout-induced $T_1$ degradation. We analytically derive the qubit decay rate from the generalized qubit spectrum under readout, including the effects of qubit-state-dependent resonator loss rates due to the presence of a Purcell filter. In particular, we show that the qubit $T_1$ can have a strong dependence on the readout drive frequency $\omega_d$ even when the  measurement rate is held fixed.  We find that driving at $\omega_d = \omega_r(g)$ (the resonator frequency when the qubit is in the ground state) yields the narrowest spectral linewidth [see Fig.~\ref{fig:cartoon}(d)], and consequently the least $T_1$ degradation in the presence of a detuned TLS. We validate our theory and this prediction through both numerical simulations and experimental measurements on a superconducting transmon qubit, finding excellent agreement.  Notably, these results cannot be explained by the simple Lorentzian broadening model.

Our paper is structured as follows. In Sec.~II, we derive a rate equation for predicting readout-induced qubit $T_1$ suppression and present numerical simulations confirming its validity. In Sec.~III, we further validate our theory by comparing measured qubit decay rates against the theoretical predictions. Finally, Sec.~IV summarizes our findings and discusses their implications and outlook.

\section{Theory for lifetime degradation}

\subsection{Deriving $T_1$ with renormalized qubit spectrum}

For the $T_1$ degradation mechanism we seek to model here, it is sufficient to work with a purely dispersive interaction between the qubit and resonator, and neglect higher transmon levels. In a frame rotating at the drive frequency, the Hamiltonian for the coupled qubit and resonator is \cite{koch_2007, Blais_CircuitQED_Review}
\begin{align} \tilde{H}_{qr} =&\, \omega_q\hat{\sigma}_+\hat{\sigma}_- - \Delta_d \hat{a}^\dagger\hat{a} + \chi\hat{\sigma}_+\hat{\sigma}_-\hat{a}^\dagger\hat{a}\nonumber\\ 
&+ d_r\Big(1+\frac{\delta_p}{2}\hat{\sigma}_z\Big)(\hat{a}^\dagger + \hat{a}).\label{eq:Hamiltonian_rot} \end{align}
Here, $\hat{\sigma}_-$ ($\hat{\sigma}_+$) denotes the lowering (raising) operator of the qubit, $\hat{\sigma}_z=[\hat{\sigma}_+,\hat{\sigma}_-]$ is the Pauli-$z$ operator, and $\hat{a}$ ($\hat{a}^\dagger$) denotes the annihilation (creation) operator of the resonator. The frequencies to excite the qubit and resonator when the hybrid system is in its ground state are $\omega_q$ and $\omega_r(g) \equiv \omega_r$, respectively. The drive detuning is given by $\Delta_d = \omega_d - \omega_r(g)$, where $\omega_d$ is the lab-frame readout drive frequency. 
Note that we define the drive detuning with respect to the cavity frequency associated with the qubit ground state; with this convention, the choice $\Delta_d = 0$ will play a special role for the problem studied in this work (readout-enhanced qubit decay).
The coefficient $\chi$ denotes the full dispersive shift, so the readout resonator frequency with the qubit in the excited state is $\omega_r(e) = \omega_r + \chi$. The drive amplitude is denoted by $d_r$, and the coefficient $\delta_p$ captures the qubit-state-dependent drive amplitude arising from the Purcell filter through which the readout drive is routed \cite{sete_readout, Swiadek_2024}. In the following, we use the shorthand notation $d_{r,e} = (1+\delta_p/2) d_r$ and $d_{r,g} = (1-\delta_p/2) d_r$.

In addition, both the resonator and qubit are coupled to independent dissipative environments. The resonator is assumed to be much lossier than the qubit. Such dissipation is generally assumed Markovian for superconducting LC oscillators, with the loss rates denoted by $\kappa$. When the Purcell filter plays a non-negligible role, one should further account for qubit-state-dependent resonator loss rates. Specifically, the damping operator in the resonator dissipator  $\kappa\,\mathbb{D}[\hat{h}]$ should take on a more general form  $\hat{h} = 
\Big(1+\delta_p\hat{\sigma}_z/2\Big)\hat{a}$, which implies that the state-dependent loss rates are $\kappa_e = \kappa (1+\delta_p/2)^2$ for the qubit excited state and $\kappa_g = \kappa (1-\delta_p/2)^2$ for the ground state~\cite{Swiadek_2024}. In the rotating frame introduced above, the steady-state resonator displacement is $\alpha_g = d_{r,g}/(\Delta_d+i\kappa_g/2)$ if the qubit is maintained in the ground state and $\alpha_e=d_{r,e}/(\Delta_d-\chi+i\kappa_e/2)$ for the excited state. Their sepration is denoted by $\delta\alpha = \alpha_e-\alpha_g$.

The qubit bath is however considered non-Markovian -- instead of being uniformly coupled to an ensemble of environmental modes, the transmon qubits are generally affected predominantly by a few discrete degrees of freedom, usually considered TLSs. We assume that the qubit–bath interaction takes the form $\hat{H}_I = \hat{\sigma}_x\hat{X}_B$, where $\hat{\sigma}_x$ is the qubit Pauli-$x$ operator and $\hat{X}_B$ is a bath operator that couples the TLSs to the qubit. We can formally write down the TLS spectrum as $S_B(\omega)=\int_{-\infty}^{\infty}dt \langle \hat{X}_B(t)\hat{X}_B(0)\rangle e^{i\omega t}$, as illustrated by the gray shapes in Fig.~\ref{fig:cartoon}(b)-(d). Here, $\hat{X}_B(t) = e^{i\hat{H}_Bt}\hat{X}_Be^{-i\hat{H}_Bt}$ is the bath operator in the interaction picture and $\hat{H}_B$ is the bare bath Hamiltonian.

We assume that the qubit-bath coupling is sufficiently weak and the qubit decay timescale is much longer than the correlation time of the resonator and qubit bath. In this way, the qubit decay rate can be calculated via time-dependent perturbation theory, resulting in an expression that is a generalized form of Fermi's golden rule 
(FGR) \cite{IBM_readout_zeno, Kofman_Kurizki}. From this we derive the qubit decay rate (see Appendix \ref{sec:corr} and \ref{sec:spectrum})
\begin{align}
    \Gamma_{e\rightarrow g}\xrightarrow{\mathrm{FGR\ limit}}  \int_{-\infty}^{\infty}\frac{d\omega}{2\pi} S_q(\omega) S_B (\omega), \label{eq:rate}
\end{align}
where the qubit emission spectrum $S_q(\omega)=\int_{-\infty}^{\infty} dt \,e^{i\omega t} C^*_q(t)$ is the Fourier transform of the two-point correlation function
\begin{align}
    C_q(t) = \langle \hat{\sigma}_+(t)\hat{\sigma}_-(0)\rangle_{e,\alpha_e}.\label{eq:corr_first}
\end{align}
The above expectation is evaluated for the initial state $\vert e,\alpha_e\rangle$ of the qubit–resonator system. This choice tacitly assumes that the qubit energy decay timescale is much longer than $1/\kappa$, so that the resonator reaches its steady state well before a qubit state transition occurs. In deriving the rate, we assume that the stationary condition is satisfied for both the bath and qubit correlation functions, on the basis that both the bath relaxation and resonator decay times are much shorter than the $T_1$ timescale of interest.

\begin{figure*}
    \centering
    \includegraphics[width=0.84\linewidth]{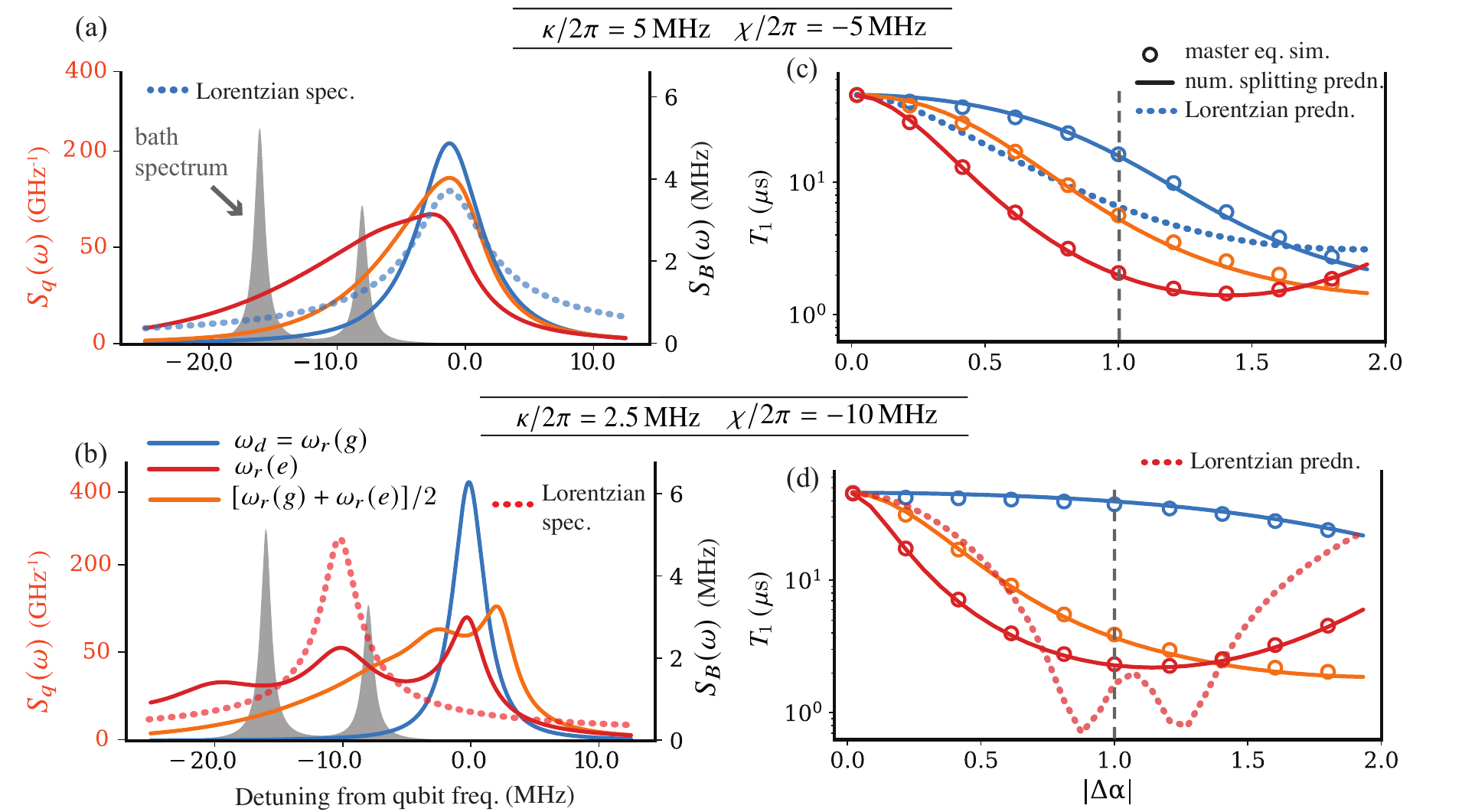}
\caption{Numerical verification of the analytical model. We consider a transmon qubit coupled to two TLSs and dispersively coupled to a driven, lossy resonator. (a,b) Qubit emission spectra $S_q(\omega)$ under readout drive for (a) $\chi=\kappa$ and (b) $\chi=4\kappa$. Different colors correspond to different readout frequencies: $\omega_d = \omega_r(g)$ (blue), $\omega_d = \omega_r(e)$ (red), and $\omega_d = [\omega_r(g) + \omega_r(e)]/2$ (orange). Gray shading indicates the TLS noise spectrum, with TLS-transmon coupling strengths set as $g_{\mathrm{tls}}/2\pi = 0.5$ MHz (left) and $0.4$ MHz (right) and loss rates $\gamma_{2,\mathrm{tls}}/2\pi = 0.5$ MHz. (c,d) Qubit decay rate $\Gamma_{e \to g}$ as a function of pointer-state separation $|\delta\alpha|$. Solid lines: predictions from Eqs.~\eqref{eq:rate} and \eqref{eq:spectrum}; open circles: master equation simulation results. Dashed lines: predictions using the simple Lorentzian broadening model for (c) $\omega_d = \omega_r(g)$ and (d) $\omega_d = \omega_r(e)$, which deviate significantly from simulations. The excellent agreement between solid lines and markers validates our analytical model across both coupling regimes.}
    \label{fig:numerics}
\end{figure*}
 
The correlation function \eqref{eq:corr_first} can be analytically derived using the quantum regression theorem, as in Refs.~\cite{Gambetta_num_split_theor,Clerk_num_split}. Here we generalized the results there to include state-dependent drive amplitudes and decay rates. The evolution of the qubit operator in $C_q(t)$ is governed by the propagator $e^{\mathcal{L}t}$, where  the Lindbladian is given by $\mathcal{L} = -i [\tilde{H}_{qr}, \cdot] + \kappa\,\mathbb{D}[\hat{h}]$. The resulting correlation function for the \textit{emission} spectrum is given by
\begin{align}
    C_q(t) = &\exp[i(\omega_q+B)t-\Gamma_m t] \nonumber\\&\times\exp\Big\{  
    A \Big[1-\exp[(i\Delta_d-\kappa_g/2)t]\Big]  \Big\},\label{eq:corr}
\end{align}
which further leads to the qubit spectrum 
\begin{align}
    S_q(\omega) = \sum_{j\in\mathbb{N}} \frac{2}{j!}\mathrm{Re}\Bigg[\frac{(-A)^je^A}{\Gamma_m+j\kappa_g/2 - i(\omega-\omega_j)}\Bigg]. \label{eq:spectrum}
\end{align}
Above, $B$ is the drive induced overall frequency shift, and $\Gamma_m$ is the measurement induced dephasing rate given by $\Gamma_m = |\sqrt{\kappa_e}\alpha_e-\sqrt{\kappa}_g\alpha_g|^2/2$, which reduces to $\kappa|\delta\alpha|^2/2$ for symmetrical loss rates. The spectrum \eqref{eq:spectrum} is a weighted infinite sum of poles, with the weights decaying  exponentially with $j$ as $\sim (-A)^j$ where $A$ is a complex number which in most cases has the magnitude $|\delta \alpha|^2$. The expressions for $A$ and $B$ can be found in Appendix \ref{sec:spectrum}. The center frequencies for these poles are $\omega_j = \omega_q + j\Delta_d + B$, whose separation is determined by the drive detuning $\Delta_d$. Note that the above expressions assume a zero-temperature cavity bath and should be modified for finite temperature \cite{Clerk_num_split}.

Several features of the qubit emission spectrum $S_q(\omega)$ can be understood directly from the analytical expressions. When $|A| \ll 1$, or equivalently $|\delta\alpha|^2 \ll 1$, the second exponential in Eq.~\eqref{eq:corr} contributes negligibly, and $S_q(\omega)$ reduces to a single Lorentzian centered at $\omega_q + B$ with width $\Gamma_m$. This is the regime in which the conventional Lorentzian broadening model applies. For larger $|\delta\alpha|^2$, however, higher-order terms in the sum become significant and the spectral shape is governed by the pole structure of Eq.~\eqref{eq:spectrum}. The center frequencies of these poles are $\omega_j = \omega_q + j\Delta_d + B$, whose separation is determined by the drive detuning $\Delta_d$
[defined with respect to $\omega_r(g)$]. This immediately suggests that the choice of readout drive frequency shapes the width of the qubit emission spectrum. In particular, driving at $\omega_d = \omega_r{(g)}$ ($\Delta_d = 0$) causes all poles to coincide, minimizing the overall spectral width.

Conversely, when $\Delta_d$ is non-zero — for example, $\Delta_d = \chi$ corresponding to driving at $\omega_d = \omega_r{(e)}$ — the poles are separated, and become resolved for $\kappa_g < |\Delta_d|$, as shown by the red curve Fig.~\ref{fig:cartoon}(d).  In this way, the spectrum exhibits discrete peaks, a hallmark of the number-splitting regime. 

The origin of these discrete peaks can be understood as a multi-wave mixing process, which becomes apparent after performing a polaron transformation~\cite{Gambetta_trajectory} (see Appendix~\ref{sec:multi-wave-mixing} for details). The key insight is that the qubit lowering operator transforms as $\hat{\sigma}_- \rightarrow \hat{\sigma}_-\exp(\delta\alpha\hat{a}^\dagger - \delta\alpha^*\hat{a})$. As a result, a qubit decay event, i.e., the conversion of a qubit excitation to a bath excitation, can be accompanied by the creation of photons in the resonator. Such creation necessarily involves the absorption of a drive photon-- drive photons are converted to cavity photons during the qubit relaxation event. Energy conservation then requires 
\begin{align}
    \omega_B=  \tilde{\omega}_q + n[\omega_d - \omega_r(g)] ,
\end{align}
where $\omega_B$ is the bath photon frequency, $\tilde{\omega}_q\equiv\omega_q+B$ is the drive-shifted qubit frequency, and $n$ is the net number of photons added to the resonator. Each value of $n$ corresponds to a distinct peak in the qubit emission spectrum, with their separation determined by the drive detuning $\Delta_d = \omega_d - \omega_r(g)$. Note crucially that the energy conservation condition involves $\omega_r(g)$, the cavity frequency associated with the qubit ground state.  This reflects the simple fact that the final state of these relaxation processes have the qubit in its ground state, hence any photons produced have an energy $\omega_r(g)$.  We immediately see that the choice $\Delta_d = 0$ is special: in this case, conversion of drive photons to cavity photons costs no energy, and all of the above multi-photon processes coincide, i.e.~$\omega_B$ becomes independent of $n$.

The lowest-order nontrivial process is illustrated as a four-wave mixing diagram in Fig.~\ref{fig:cartoon}(d) inset. This photon creation mechanism implies that a qubit energy decay event during measurement may cause the resonator field to deviate from a coherent state, potentially introducing an additional source of readout error that merits further study. We note that although Eq.~\eqref{eq:spectrum} has a discrete-peak structure for $\omega_d = \omega_r{(e)}$ and $\kappa_g\ll \chi$, in the large-$|A|$ limit the Poisson weighting $(-A)^j e^A/j!$ can mimic a broad, smooth envelope. Specifically, since $|A| \approx \bar{n}_e$, the weights are concentrated around $j \approx \bar{n}_e$ with width $\sim \sqrt{\bar{n}_e}$, where $\bar{n}_e$ is the mean photon number when the qubit is in $|e\rangle$. The resulting spectrum thus approximates a Poisson envelope centered at $\omega_q + \chi \bar{n}_e$.

Finally, we comment on the apparent asymmetry between $\omega_d = \omega_r{(g)}$ and $\omega_d = \omega_r{(e)}$ in linewdith broadening. This asymmetry arises because $S_q(\omega)$ describes a qubit \emph{emission} process. For the qubit \emph{absorption} spectrum as focused on in Refs.~\cite{Gambetta_num_split_theor}, the roles of $\omega_r{(g)}$ and $\omega_r{(e)}$ would be reversed. A more detailed physical discussion is provided in Appendix~\ref{sec:multi-wave-mixing}.

\subsection{Numerical simulation}

The expression Eq.~\eqref{eq:rate} is compact and physically transparent, but the validity of the Fermi's golden rule treatment relies on the separation of timescales between resonator relaxation, bath relaxation versus qubit decay. To verify that this approximation remain valid across the parameter regimes of interest, we perform master equation simulations of the full model described in the previous subsection \cite{johansson2012qutip}, with TLSs included as explicit quantum modes. 

The resonator-qubit system Hamiltonian is given in Eq.~\eqref{eq:Hamiltonian_rot}. The TLS configurations are illustrated in Fig.~\ref{fig:numerics}(a) and (b), where the gray shading indicates the TLS noise spectrum with detunings visible relative to the qubit frequency $\omega_q$. Their parameters are given in the figure caption. We consider two representative $|\chi|/\kappa$ regimes: a conventional choice with $\chi/2\pi = -5$ MHz and $\kappa/2\pi = 5$ MHz [Fig.~\ref{fig:numerics}(a, c)], and a more radical configuration with $\chi/2\pi = -10$ MHz and $\kappa/2\pi = 2.5$ MHz [Fig.~\ref{fig:numerics}(b, d)], corresponding to $|\chi|/\kappa = 1$ and $|\chi|/\kappa = 4$. In these simulations, we neglect the Purcell-filter-induced dependence of drive strengths and decay rates on qubit states (i.e., we set $\delta_p = 0$) to focus on the most fundamental effects of the qubit spectral reshaping. The master equation results are further compared to predictions from Eq.~\eqref{eq:rate}, where the bath spectrum $S_B(\omega)$ is derived using the standard TLS theory \cite{tls_spectrum,you_tls_th}. To ensure a fair comparison with our analytical predictions, the numerical simulations are performed with the system initialized in the steady state $\vert e, \alpha_e\rangle$. We examine in particular the $T_1$ degradation as a function of readout drive frequency at fixed $\Gamma_m$, which is also the ideal steady-state measurement rate.

Even in the conventional regime $|\chi|/\kappa = 1$ , the qubit spectrum already exhibits non-trivial features. Fig.~\ref{fig:numerics}(a) shows the qubit emission spectra in this regime for a fixed pointer-state separation $|\delta\alpha| = 1$ and three different readout drive frequencies: $\omega_d = \omega_r(g)$ (blue), $\omega_d = \omega_r(e)$ (red), and $\omega_d = [\omega_r(g) + \omega_r(e)]/2$ (orange). Despite identical $\Gamma_m$, the spectral shapes differ substantially depending on the choice of $\omega_d$. As we stated in the last subsection, this difference originates from the role of drive detuning $\Delta_d$ in determining the peak separation in Eq.~\eqref{eq:rate}. Consequently, the three spectra overlap differently with the TLS bath spectrum (gray shading), yielding distinct decay rate predictions according to Eq.~\eqref{eq:rate}: driving at $\omega_r(g)$ results in the least $T_1$ degradation, while driving at $\omega_r(e)$ leads to the most significant lifetime reduction. As shown in Fig.~\ref{fig:numerics}(c), the master equation simulation results (markers) agree excellently with our number-splitting predictions (solid lines).

A particularly striking example for comparing the number-splitting and Lorentzian prediction is $\omega_d = \omega_r(g)$. As shown in Fig.~\ref{fig:numerics}(a), the former (blue solid) is significantly narrower than the latter (dashed) — and master equation simulations confirm that the number-splitting result is the accurate one. Consequently, the Lorentzian model substantially overestimates the qubit decay rate [Fig.~\ref{fig:numerics}(c)]. This narrowing can be understood physically from the correlation function~\eqref{eq:corr}. At short times, the qubit dephasing due to photon-number fluctuation must grow quadratically in $t$, since the resonator has a finite correlation time $\sim 1/\kappa$. This quadratic onset, in contrast to the linear-in-$t$ dephasing that produces a Lorentzian, leads to a spectrum that decays like a Gaussian at large frequencies—much faster than a Lorentzian tail.

For the larger $|\chi|/\kappa = 4$ ratio, the deviation from the Lorentzian model becomes even more pronounced. As shown in Fig.~\ref{fig:numerics}(b), the qubit spectrum exhibits qualitatively different behavior depending on the choice of $\omega_d$. For $\omega_d = \omega_r(g)$ (blue), the spectrum remains a single peak; for $\omega_d = \omega_r(e)$ (red) and the center frequency (orange), the spectrum exhibits multiple resolved peaks — a signature of number splitting. Taking $\omega_d = \omega_r(e)$ as an example, the emergence of these discrete peaks leads to a qubit decay rate that differs substantially from what one would predict by assuming a simple Lorentzian-broadened, ac-Stark-shifted spectrum [dashed line in Fig. \ref{fig:numerics}(d)].  
Fig.~\ref{fig:numerics}(b) also shows another surprising feature:
for the standard choice of drive detuning $\omega_d = [\omega_r(e)+\omega_r(g)]/2$ (orange curve), the dominant rightmost peak in the qubit spectrum has a {\it positive} shift with respect to the bare qubit frequency.  This corresponds to the  non-trivial behavior of the qubit AC Stark shift at large $\chi$, as given by the constant $B$: for this detuning $B$ changes sign when $\chi$ is increased past $\kappa/2$.  We provide more details on this phenomenon in Appendix \ref{sec:multi-wave-mixing} via the polaron-frame description of the dynamics \cite{Gambetta_trajectory}. The excellent agreement between our analytical prediction and the master equation results (markers) across both regimes validates the necessity of incorporating number-splitting physics when modeling readout-induced $T_1$ degradation.

\section{Experimental verification}
\begin{figure*}
    \centering
    \includegraphics[width=0.9\linewidth]{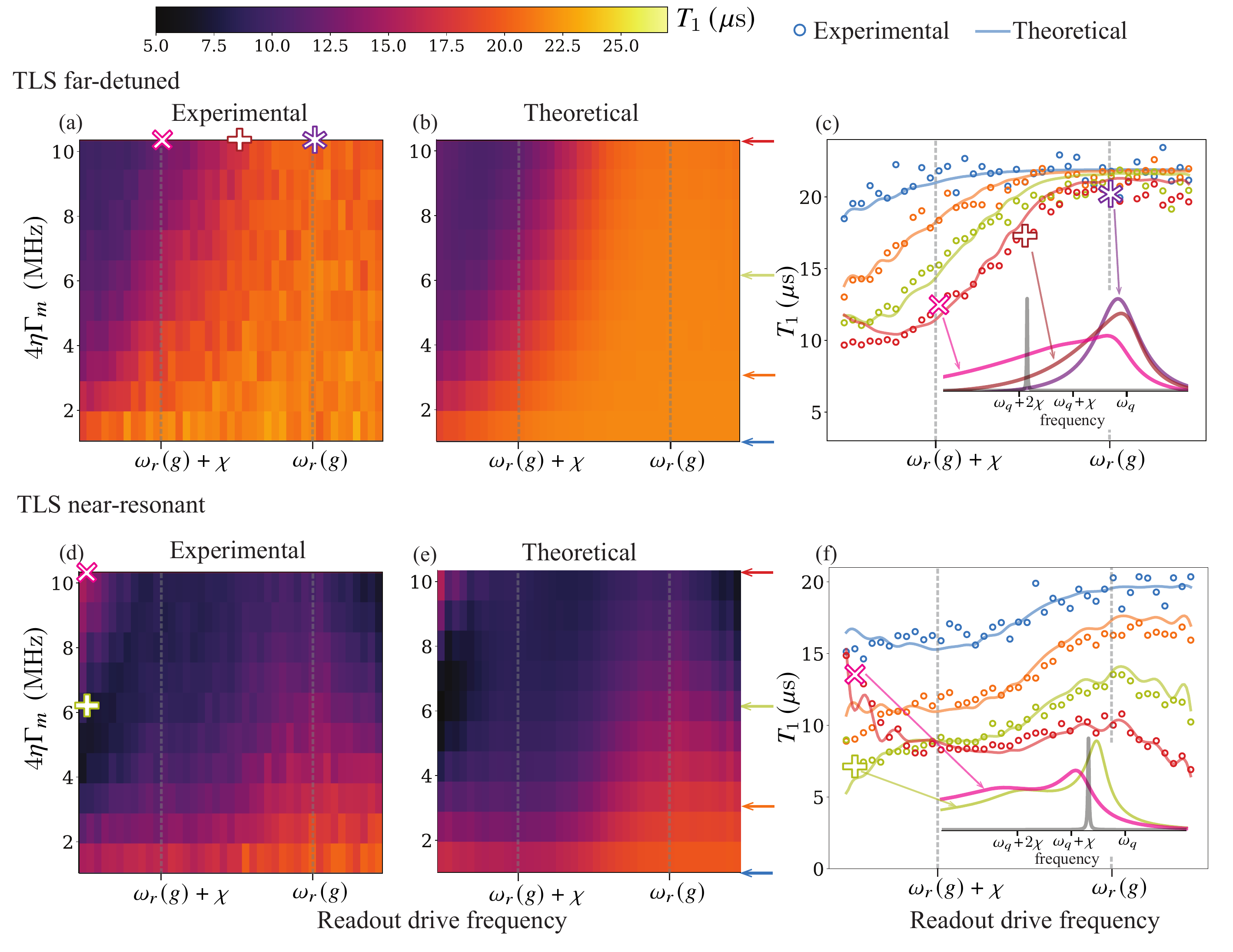}
\caption{Quantitative comparison between experimental data and theoretical calculation.
{(a-c)} Data for a TLS far-detuned from the qubit frequency ($\Delta_{\mathrm{tls}}\approx 2\chi$). 
{(d-f)} Data for a TLS near-resonant with the qubit frequency ($\Delta_{\mathrm{tls}}\lesssim \chi$). 
{(a, d)} Experimental qubit relaxation time $T_1$ (colorscale, in $\mu$s) as a function of 
readout drive frequency (horizontal axis) and SNR rate (vertical axis, in MHz).  The standard errors of the extracted $T_1$ values are small across all data sets (median 3\%). The TLS has $g_{\mathrm{tls}}/2\pi\approx 0.2$ MHz and $g_{\mathrm{tls}}/\gamma_{2,\mathrm{tls}}= 1.4\,(0.9)$ for the top (bottom) TLS configuration (see details in Appendix \ref{app:tls}). Above, we denote the coupling strength $g_\mathrm{tls}$, 
detuning $\Delta_\mathrm{tls} \equiv \omega_\mathrm{tls} - \omega_q$, and 
TLS dephasing rate $\gamma_{2,\mathrm{tls}}$. The drive 
frequency is swept across the $g$ state resonant frequency $\omega_r(g)$ and $e$ state frequency $\omega_r{(g)} + \chi$, where $\omega_r{(g)}$ 
is the bare resonator frequency and $\chi$ is the dispersive shift. 
{(b, e)} Corresponding theoretical calculations using the emission-spectrum formalism, 
reproducing the key features observed in the experimental data including the $T_1$ suppression 
near specific readout drive frequencies. 
{(c, f)} Horizontal linecuts comparing experimental data (open circles) and theoretical 
predictions (solid lines) at SNR rates indicated by the colored arrows in (b) and (e), 
demonstrating quantitative agreement across the full frequency range. 
Insets in (c) and (f) display the TLS spectral 
density $S_{B}(\omega)$ (gray) and emission spectra $S_q(\omega)$ (other coloring) for representative operating points labeled by 
the same markers in (a, d) and (c, f). The degree of spectral overlap between the qubit emission spectra and TLS spectrum directly determines 
the qubit relaxation rate: larger overlap leads to enhanced $T_1$ degradation.}
    \label{fig:data}
\end{figure*}

The model for readout-induced interaction with the TLS bath is tested experimentally using a frequency-tunable transmon qubit coupled strongly to a Purcell-filtered readout resonator. The device has $\chi/2\pi = -8.8$ MHz, with qubit-state-dependent resonator loss rates $\kappa_g/2\pi = 9.0$ MHz and $\kappa_e/2\pi = 6.6$ MHz. This considerable difference between $\kappa_g$ and $\kappa_e$ arises from the large dispersive shift, which modifies the detuning between the readout resonator and the Purcell filter depending on the qubit state~\cite{Swiadek_2024}. Our device thus provides a concrete example where the qubit-state-dependent loss rates are non-negligible, making the generalized model in Eq.~\eqref{eq:Hamiltonian_rot} necessary for an accurate description of the readout-induced dynamics.

We leverage the flux tunability of the transmon to perform two sets of experiments: with the readout drive off, we sweep the qubit frequency across the TLS bath to characterize frequency-dependent qubit decay rates and the coupling strengths of individual strongly-coupled TLSs \cite{Barends_2013}; with the qubit frequency fixed and detuned from the TLS, we apply a continuous-wave readout drive and measure the qubit decay rate, probing the readout-induced interaction with the TLS as predicted by our theory. In both cases, we obtain qubit decay traces from inversion recovery experiments, in which the qubit is prepared in the excited state and its population is monitored as a function of delay time. The experiments with and without readout drives are interleaved tightly such that temporal reconfigurations of the TLS bath can be better resolved. These fluctuations, which can occur on the scale of minutes, are captured by analyzing continuous shots of the experiment where each shot consisting of the sweep over all parameters takes $\sim$2 seconds.

To explore the drive frequency dependence of the qubit's interaction with the bath, we measure the qubit decay rates under readout over a sweep of readout frequencies while holding constant experimental measurement rates $\frac{d}{dt} \text{SNR} = 4\eta\Gamma_m$ where $\eta$ is the quantum efficiency and $\Gamma_m$ is the readout-induced dephasing rate. The readout drive strength at the device is also calibrated for theoretical modeling of the experiment by characterizing the ac-Stark shift of the qubit frequency when subjected to the readout drive given the measured $\chi$ \cite{Sank_Ross}. See Appendix \ref{app:exp-system} for details of calibrations and experimental parameters.
We estimate that at most $\bar{n} \approx 3$ photons reside in the resonator across all readout settings used here, while semiclassical Floquet simulations of the critical photon number indicate no significant MIST channels below this level \cite{BlaisIonization}. Consistently, the single-shot readout signals show no significant leakage state distributions, and we bound the leakage population out of the $\{|g\rangle, |e\rangle\}$ manifold to below $1\%$, ruling out MIST as a likely cause of the observed $T_1$ reduction.

To test our model for $T_1$ degradation experimentally, we require knowledge of the bath noise spectral density $S_B(\omega)$. For most qubit frequencies, we can fit the measured decay trace using a single exponential to extract the decay rate, which according to Fermi's golden rule is equivalent to $S_B(\omega)$. While such extraction is straightforward, the fitted decay rates fluctuate considerably over timescales of minutes, making direct use of the resulting $S_B(\omega)$ less reliable. Near resonance with a strongly and coherently coupled TLS, the qubit decay trace deviates from a single exponential. The coherent qubit–TLS swap dynamics imprint a characteristic double-exponential envelope on the decay trace, from which the coupling strength $g_\mathrm{tls}$ and its dephasing rate $\gamma_{2,\mathrm{tls}}$ can be extracted \cite{muller2009} (see Appendix \ref{app:tls} for the full functional form). With these two parameters carefully calibrated, one can calculate the defect's contribution to $S_B(\omega)$ from standard TLS theory~\cite{tls_spectrum,you_tls_th}.

Motivated by these considerations, we focus our experiment–theory comparison on a single, stable, strongly coupled TLS, which simplifies the comparison: its contribution to $S_B(\omega)$ can be accurately characterized and overwhelms the unstable but weak qubit-decay background. We note, however, that the theoretical model readily extends to complex bath spectra representing many more degrees of freedom. In our measurement, we identify a frequency band containing a single prominently coupled TLS rather than a region with many. During the experiment, this specific TLS switches infrequently—on the timescale of tens of minutes—between two distinct frequency configurations (see Appendix~\ref{app:tls}). We accordingly segment the interleaved idle-$T_1$ and $T_1$-under-readout data into two cases exhibiting distinct readout-induced relaxation behaviors: one in which the TLS is far-detuned from the qubit frequency with $\Delta_{\text{tls}}/2\pi \approx -16$ MHz [Figs.~\ref{fig:data}(a–c)], and another in which it is closer to resonance with $\Delta_{\text{tls}}/2\pi \approx -6$ MHz [Figs.~\ref{fig:data}(d–f)]. Here, $\Delta_{\mathrm{tls}}$ denotes the detuning between the TLS and the qubit idle frequencies.

Fig.~\ref{fig:data} presents experimental data taken across a range of readout drive frequencies and measurement rates alongside the corresponding predictions of our theoretical framework.  We first examine the far-detuned TLS case shown in Figs.~\ref{fig:data}(a–c). The experimental data in Fig.~\ref{fig:data}(a) display the measured qubit relaxation time $T_1$ as a function of readout drive frequency (horizontal axis) and $\frac{d}{dt}\mathrm{SNR}$ (vertical axis), with the drive frequency swept across the range from the ground-state resonator frequency $\omega_r{(g)}$ to the excited-state frequency $\omega_r{(g)} + \chi$. The corresponding theoretical calculation, shown in Fig.~\ref{fig:data}(b), exhibits excellent agreement with the experimental data — not only do the overall features match, but the horizontal linecuts in Fig.~\ref{fig:data}(c) demonstrate quantitative agreement across the full frequency range at multiple SNR rate. We emphasize that this agreement is achieved using the generalized model with $\delta_p 
\neq 0$, which properly captures the qubit-state-dependent drive amplitudes and loss rates inherited from the Purcell-filter coupling.

A central observation from this dataset is that, even at fixed $\frac{d}{dt}\mathrm{SNR}$, the measured $T_1$ varies significantly with the choice of readout frequency. This behavior directly reflects the frequency-dependent spectral broadening discussed in Sec. II: different choices of $\omega_d$ yield qubit emission spectra with distinct shapes and widths, leading to different degrees of overlap with the TLS bath spectrum. To illustrate this point, the inset of Fig.~\ref{fig:data}(c) displays the qubit emission spectrum $S_q(\omega)$ for three representative readout frequencies at the maximum SNR rate, alongside the TLS spectral density $S_B(\omega)$ (gray shading). Despite identical measurement rates, the three qubit emission spectra differ markedly in their spectral extent and, consequently, in their overlap with the TLS spectrum. Driving at $\omega_r{(g)}$ produces the narrowest spectrum with minimal TLS overlap, while driving at $\omega_r{(e)}$ yields a broader spectrum that extends further toward the TLS resonance. The difference in the overlap directly accounts for the observed variation in $T_1$ across readout frequencies.

The near-resonant TLS case, shown in Figs.~\ref{fig:data}(d–f) with $\Delta_{\text{tls}} \approx -6$ MHz, offers richer physics due to the proximity between the TLS and the qubit frequency. As before, the theoretical predictions in Fig.~\ref{fig:data}(e) show excellent agreement with the experimental data in Fig.~\ref{fig:data}(d), and the linecuts in Fig.~\ref{fig:data}(f) confirm this agreement quantitatively, which further validates our generalized number-splitting framework as a robust and predictive tool. Crucially, because the TLS now lies within the spectral range accessible to the broadened qubit emission spectrum, increasing the readout drive extends rather than shortens $T_1$. This is a hallmark of the quantum Zeno effect \cite{IBM_readout_zeno}, in which stronger measurement suppresses decay. 

The highlighted markers in Figs.~\ref{fig:data}(d–f) illustrate this effect: these points correspond to measurements taken at fixed readout frequency but varying drive amplitudes. Counterintuitively, $T_1$ increases at stronger drive powers despite the nominally more invasive measurement. This revival can be understood through the spectral overlap picture: as the drive amplitude increases, the qubit emission spectrum becomes increasingly diluted across multiple number-split peaks, reducing the spectral weight at any single frequency. For a near-resonant TLS, this dilution decreases the overlap between $S_q(\omega)$ and $S_B(\omega)$, leading to a suppressed decay rate according to Eq. (3). The inset of Fig.~\ref{fig:data}(f) illustrates this mechanism, showing how the qubit emission spectrum evolves from a concentrated single peak at low drive power to a broader, multi-peaked structure at high drive power with progressively less overlap with the TLS resonance.

\section{Conclusion and outlook}

In this work, we established a theoretical framework for predicting readout-induced $T_1$ degradation that fully incorporates number-splitting physics, moving beyond the conventional Lorentzian broadening model. We demonstrated that for modest to strong $|\chi|/\kappa$ ratios, the qubit emission spectrum exhibits clearly non-Lorentzian broadening, fundamentally altering the qubit's interaction with environmental TLSs. Both master equation simulations and experimental measurements show excellent agreement with our analytical predictions, validating the framework across a range of drive conditions.

A central finding of this work is that the readout drive frequency $\omega_d$ provides a previously unrecognized degree of freedom for controlling TLS-induced lifetime degradation. We show rigorously, through both theory and experiment, that for the same measurement rate, driving at $\omega_r{(g)}$ yields a substantially narrower qubit emission spectrum than driving at $\omega_r{(e)}$, significantly reducing the qubit's sensitivity to background TLSs. This result has direct implications for superconducting quantum processors operating in the strong dispersive regime, where careful selection of readout frequency can mitigate TLS-mediated relaxation during measurement without sacrificing readout performance. The underlying mechanism, namely that readout-induced spectral broadening driving unwanted transitions, extends beyond TLS defects. Recent work has shown that readout photons can similarly drive crosstalk with spectator qubits \cite{Readout_crosstalk}. Our findings suggest that readout frequency optimization may serve as a complementary mitigation strategy for these correlated errors in multi-qubit processors.

Our framework also opens new directions for future research. The frequency-resolved and width-tunable nature of the readout-induced emission spectrum offers a novel method for characterizing the background TLS bath, potentially circumventing the frequency resolution limitations inherent in conventional $T_1$ spectroscopy. Furthermore, our analysis reveals the possibility of optimizing readout pulse shapes and protocols to further reduce qubit sensitivity to environmental defects. These results provide a foundation for predicting and minimizing readout-related errors in next-generation quantum error correction architectures, where fast, high-fidelity readout must be balanced against environmental decoherence mechanisms.

\begin{acknowledgments}
We thank the staff from across the Amazon Center for Quantum Computing that enabled this project.
\end{acknowledgments}

\appendix

\section{Qubit correlation functions under readout considering $g$, $e$ asymmetry}
\label{sec:corr}

For self-containment, here we show derivation of the correlation function $ C_q(t) = \langle \hat{\sigma}_+(t)\hat{\sigma}_-(0)\rangle_{e,\alpha_e}$ using the quantum regression theorem, which is 
\begin{align}
    C_q(t) = \mathrm{Tr}\Big\{\hat{\sigma}_+ e^{\mathcal{L}t}\vert g,\alpha_e\rangle\langle e,\alpha_e\vert\Big\}.
\end{align}
Above, the Lindbladian governing the qubit-resonator state evolution is $\mathcal{L} = -i [\tilde{H}_{qr}, \cdot] + \kappa\,\mathbb{D}[\hat{h}]$, where we consider a more general damping operator $\hat{h} = (x\vert e\rangle\langle e\vert + y\vert g\rangle\langle g\vert)\hat{a}$ and a drive $\tilde{H}_d(t) = d(a^\dagger+a)(u\vert e\rangle\langle e\vert + v\vert g\rangle\langle g\vert)$. The damping and drive operator reduce to the standard forms by setting ${x}=y=1$ and $u=v=1$.

In this and many other experimental settings, the readout resonator is coupled to the drive line through a Purcell filter \cite{sete_readout,Swiadek_2024}. As a result, both the readout drive strength and the loss rate depend on the hybridization between the readout resonator and the Purcell filter, which in turn depends on the qubit state, primarily through the dispersive shift. This dependence breaks the equalities $x = y$ and $u = v$ \cite{Swiadek_2024}. More specifically, the annihilation operator of the bare Purcell-filter mode, denoted by $\hat{p}$, transforms into $\hat{p}\rightarrow\hat{p}\cos{\theta}^{\nu}_{rp} + \hat{a}\sin{\theta}^{\nu}_{rp}$ after the hybridization, where the hybridization angle $\theta^\nu_{rp}$ depends on the qubit state $\nu = g,e$. Within this model, the four dimensionless factors $x,y,u$ and $v$ can be related by $x/y=u/v = \sin\theta^e_{rp}/\sin\theta^g_{rp}$. Using the same notation, we can also properly define the asymmetry coefficient $\delta_p$ in Eq.~\eqref{eq:Hamiltonian_rot} as  $\delta_p = 2(\sin\theta^e_{rp} - \sin\theta^g_{rp})/ (\sin\theta^e_{rp}+ \sin\theta^g_{rp})$. In our following treatment, we keep all four factors $x,y,u$ and $v$ to maintain a general expression. 

For convenience, we define $\hat{\rho}_{ge}(t) \equiv e^{\mathcal{L}t}\vert g,\alpha_e\rangle\langle e,\alpha_e\vert $. The derivation of $C_q(t)$ is simplified by adopting the following ansatz:
\begin{align}
    \hat{\rho}_{ge}(t) = \frac{C_q(t)}{\langle \bar{\alpha}_e(t)\vert \bar{\alpha}_g(t)\rangle} \vert g, \bar{\alpha}_g(t)\rangle\langle e,\bar{\alpha}_e(t)\vert.\label{eq:ansatz}
\end{align}
The denominator on the right-hand side ensures consistency with the definition of $C_q(t)$. Specifically, since
\begin{align}
\mathrm{Tr}\{\hat{\sigma}_+ \vert g, \bar{\alpha}_g(t)\rangle\langle e,\bar{\alpha}_e(t)\vert\} = &\,\langle \bar{\alpha}_e(t)\vert \bar{\alpha}_g(t)\rangle\\=&\, e^{-\frac{1}{2}(|\bar{\alpha}_e|^2+|\bar{\alpha}_g|^2-2\bar{\alpha}_e^*\bar{\alpha}_g)},\nonumber
\end{align}
taking the trace $\mathrm{Tr}\{{\hat{\sigma}_+\hat{\rho}_{ge}(t)\}}$ with the ansatz yields $C_q(t)$ directly, confirming its validity.

To arrive at an equation of motion, we take time derivatives of the both sides of Eq.~\eqref{eq:ansatz}. The derivative of the left-hand side is governed by the Lindbladian, i.e.,
\begin{align}
    \frac{d}{dt}\hat{\rho}_{ge}\! = &\,i[\Delta_d \hat{a}^\dagger\hat{a}, \hat{\rho}_{ge}] -i[\tilde{H}_d, \hat{\rho}_{ge}] +i\hat{\rho}_{ge}(\chi\hat{a}^\dagger\hat{a}+\omega_q)\nonumber\\
    &+\kappa \hat{h}\hat{\rho}_{ge}\hat{h}^\dagger - \frac{1}{2}\kappa \hat{h}^\dagger\hat{h}\hat{\rho}_{ge} -\frac{1}{2}\kappa\rho_{ge} \hat{h}^\dagger\hat{h}\nonumber\\
    =\, \Big[&i\Delta_d \bar{\alpha}_g\hat{a}^\dagger - ivd(\bar{\alpha}_g +\hat{a}^\dagger) -\frac{1}{2}y^2\kappa \bar{\alpha}_g a^\dagger \Big]\hat{\rho}_{ge}\nonumber\\
    +&\hat{\rho}_{ge}\Big[i(\chi-\Delta_d) \bar{\alpha}^*_e\hat{a} + iud(\bar{\alpha}^*_e +\hat{a}) -\frac{1}{2}x^2\kappa \bar{\alpha}^*_e \hat{a}\Big] \nonumber\\
    +&i\omega_q\hat{\rho}_{ge}+ xy\,\kappa\, \bar{\alpha}_g \bar{\alpha}_e^*.\label{eq:LHS}
\end{align}
The derivative of the right-hand side is evaluated by the Leibniz rule. We first calculate
\begin{align}
    \frac{d }{dt} [{\langle \bar{\alpha}_e\vert \bar{\alpha}_g\rangle}]^{-1} =&\,[{\langle \bar{\alpha}_e\vert \bar{\alpha}_g\rangle}]^{-1}\times\frac{1}{2}\Big[ \bar{\alpha}^*_e\dot{\bar{\alpha}}_e+\bar{\alpha}_g\dot{\bar{\alpha}}_g^*\nonumber\\
    & \quad+ (\bar{\alpha}_e-2\bar{\alpha}_g )\dot{\bar{\alpha}}_e^*+ (\bar{\alpha}_g^*-2\bar{\alpha}_e^* )\dot{\bar{\alpha}}_g\Big]\nonumber\\
    \frac{d}{dt} \vert g, \bar{\alpha}_g\rangle \langle e, \bar{\alpha}_e\vert=& \Big[\,\dot{\bar{\alpha}}_g (\hat{a}^\dagger-\frac{1}{2}\bar{\alpha}^*_g) - \frac{1}{2}\dot{\bar{\alpha}}^*_g \bar{\alpha}_g\Big]\vert g, \bar{\alpha}_g\rangle \langle e, \bar{\alpha}_e\vert\nonumber\\
    +&\vert g, \bar{\alpha}_g\rangle \langle e, \bar{\alpha}_e\vert\Big[\dot{\bar{\alpha}}^*_e (\hat{a}-\frac{1}{2}\bar{\alpha}_e) - \frac{1}{2}\dot{\bar{\alpha}}_e \bar{\alpha}_e^* \Big].
\end{align}
Using these expressions, we get
\begin{align}
    \frac{d}{dt}\hat{\rho}_{ge} = &\,\frac{\dot{C}_q(t)}{C_q(t)}\hat{\rho}_{ge} - \Big(\bar{\alpha}_g\dot{\bar{\alpha}}^*_e  + \bar{\alpha}_e^*\dot{\bar{\alpha}}_g\Big)\hat{\rho}_{ge}\nonumber\\
    &+\dot{\bar{\alpha}}_g\hat{a}^\dagger\hat{\rho}_{ge} + \hat{\rho}_{ge}\dot{\bar{\alpha}}^*_e\hat{a}.\label{eq:RHS}
\end{align}
Comparing Eqs.~\eqref{eq:LHS} and \eqref{eq:RHS}, we identify the following equations
\begin{align}
    \dot{\bar{\alpha}}_e =&- iud-i(\chi-\Delta_d )\bar{\alpha}_e - \frac{x^2\kappa}{2}\bar{\alpha}_e,\label{eq:EOM_e}\\
    \dot{\bar{\alpha}}_g =& -ivd+i\Delta_d \bar{\alpha}_g - \frac{y^2\kappa}{2}\bar{\alpha}_g, \label{eq:EOM_g}\\
    \dot{C}_q(t) =& \Big[xy\kappa\bar{\alpha}_g\bar{\alpha}^*_e\! +\! id(u\bar{\alpha}_e^*-v\bar{\alpha}_g)\! +\! \bar{\alpha}_g\dot{\bar{\alpha}}^*_e\!+\! \bar{\alpha}^*_e\dot{\bar{\alpha}}_g\Big] C_q(t)\nonumber\\
    =&\, \Big[\!-\!\frac{1}{2} (x-y)^2\kappa +i\chi \Big]\bar{\alpha}_g\bar{\alpha}_e^*C_q(t) \nonumber\\
    &+ id(u-v)(\bar{\alpha}_e^* +\bar{\alpha}_g)C_q(t)+ i\omega_q C_q(t).\label{eq:EOM_C}
\end{align}
In the last line Eq.~\eqref{eq:EOM_C}, we used the two equations of motion \eqref{eq:EOM_e} and \eqref{eq:EOM_g}. Finally, we arrive at the analytical expression of 
\begin{align}
    {C}_q(t)\! = \exp\!\Big\{&i\omega_q t+i\tilde{\chi}\!\int_0^{t}dt'\!\bar{\alpha}_g(t')\bar{\alpha}_e^*(t')\label{eq:full_corr}\\
    &+id(u-v)\!\int_0^t\! dt'[\bar{\alpha}_e^*(t) +\bar{\alpha}_g(t)]\Big\}C_q(0),\nonumber
\end{align}
where we use the shorthand notation $\tilde{\chi} \!=\!\! [ i(x\!-y)^2\kappa/2+\chi]$.

\section{Emission spectrum and decay rate}
\label{sec:spectrum}
For the qubit-resonator system prepared in state $\vert e,\alpha_e\rangle$, the initial condition is set as $\bar{\alpha}_e(0)=\bar{\alpha}_g(0)=\alpha_e$, and therefore $C_q(0)=1$. The evolution of $\bar{\alpha}_g(t)$ and $\bar{\alpha}_e(t)$ are then derived using Eqs.~\eqref{eq:EOM_g} and \eqref{eq:EOM_e} as
\begin{align}
    \bar{\alpha}_e(t) =&\, {\alpha}_e,\nonumber\\
    \bar{\alpha}_g(t) =&\, \delta\alpha \,e^{(i\Delta_d-\kappa_g/2)t}+\alpha_g,
\end{align}
where the steady-state solutions are
\begin{align}
    \alpha_e =& \frac{d_{r,e}}{\Delta_d-\chi + i \kappa_e/2 },\nonumber\\
    \alpha_g =& \frac{d_{r,g}}{\Delta_d + i \kappa_g/2 }.\label{eq:alpha_E,G}
\end{align}
Above, we denote the differential resonator decay rates by $\kappa_e = x^2\kappa$ and $\kappa_g=y^2\kappa$ and drive amplitudes $d_{r,e} = u d$ and $d_{r,g}= vd$. The separation between the two displacements is denoted by $\delta\alpha = \alpha_e-\alpha_g$. With the above results, we can separate the total correlation function into the steady-state and transient part as $C_q(t) = C^{\infty}_q(t)\cdot C_q^{\mathrm{tsnt}}(t)$. The steady-state part is 
\begin{align}
    C^{\infty}_q(t) =&\, \exp\big[-\Gamma_mt+i(\omega_q+B)t\big].
\end{align}
Above, the measurement rate is generalized as $\Gamma_m =|\sqrt{\kappa_e}\alpha_e -\sqrt{\kappa_g}\alpha_g|^2/2$ for qubit-state-dependent resonator decay rates, while the quantity \begin{align}
B = \mathrm{Im}\{\sqrt{\kappa_e\kappa_g}\alpha_e^*\alpha_g + i(d_{r,e}\alpha^*_e-d_{r,g}\alpha_g)\}\label{eq:B}
\end{align}
is the overall frequency shift. The transient part is given by
\begin{align}
    C^{\mathrm{tsnt}}_q(t) =&\, \exp\Big\{{A^*\Big[1-\exp{[(i\Delta_d-\kappa_g/2)t]}\Big]}\Big\}\nonumber\\
    =& \,e^{A^*} \sum_{j \in \mathbb{N}} \frac{(-A^*)^j}{j!}\exp{[(i\Delta_d-\kappa_g/2)t]},
\end{align}
where the coefficient $A^*$ is defined as 
\begin{align}
    A^* = \frac{i\alpha^*_e\delta\alpha\tilde{\chi} + i(d_{r,e}-d_{r,g})\delta\alpha}{i\Delta_d-\kappa_g/2}.
\end{align}
For $x=u$ and $y=v$, one can find $|A| = |\delta\alpha|^2$. Finally, by Fourier transforming $C_q(t)$, we arrive at the emission spectrum Eq.~\eqref{eq:spectrum}.

Having obtained the qubit emission spectrum $S_q(\omega)$, we next show derivation of the qubit decay rate due to coupling to the TLS bath \cite{Kofman_Kurizki}. The qubit-bath coupling is assumed to be much weaker than all other energy scales in the problem, so that the interaction $\hat{H}_{I} = \epsilon\hat{\sigma}_x \hat{X}_B$ can be treated perturbatively in the spirit of Fermi's golden rule. The dimensionless parameter $\epsilon$ is introduced to track perturbative order and is set to unity at the end of the calculation.

The qubit-resonator system is prepared in the steady state
$\hat{\rho}_s = |e,\alpha_e\rangle\langle e,\alpha_e|$. In the interaction picture, the density matrix for the system and bath $\hat{\rho}(t)$ evolution can be calculated via a Dyson series expansion. The zeroth-order solution is simply
$\hat{\rho}^{(0)}(t) = \hat{\rho}_s\otimes \hat{\rho}_{B,\mathrm{th}}$, where $\hat{\rho}_{B,\mathrm{th}}$ is the bath density operator in thermal equilibrium. To evaluate the qubit decay rate, we calculate the leading-order correction to the $\vert e\rangle$ state population:
\begin{align}
    \langle e\vert\mathrm{Tr}_{B,r}\{\hat{\rho}^{(2)}(t)\}\vert e\rangle
    =\, \mathrm{Re}\Big\{-2\epsilon^2 \int_0^t &dt'\!\int_0^{t'} dt''\;
     C_q(t'-t'')\nonumber\\
    & \times
    \langle \hat{X}_B(t')\hat{X}_B(t'')\rangle\Big\},
    \label{eq:dyson_2nd}
\end{align}
where $C_q(\tau) = \langle \hat{\sigma}_+(t')\,\hat{\sigma}_-(t'')\rangle_s$
is the qubit two-point correlation function derived in Appendix \ref{sec:corr}, and
the bath correlator
$\langle \hat{X}_B(t')\,\hat{X}_B(t'')\rangle$ depends only on the
time difference $t'-t''$ by assuming the stationary-bath condition. The symbol $\mathrm{Tr}_{B,r}\{\cdot\}$ denotes a partial trace over the qubit bath and resonator degrees of freedom. To evaluate this expression,
we expand both correlation functions in their spectral representations:
\begin{align}
    \langle \hat{X}_B(t')\,\hat{X}_B(t'')\rangle
    &= \int_{-\infty}^{\infty}\! \frac{d\omega}{2\pi}\; S_B(\omega)\,
    e^{-i\omega(t'-t'')}, \\
    C_q(t'-t'')
    &= \int_{-\infty}^{\infty}\! \frac{d\omega}{2\pi}\; S_q(\omega)\,
    e^{i\omega(t'-t'')},
\end{align}
where $S_B(\omega)$ is the bath noise spectral density and $S_q(\omega)$
is the qubit emission spectrum obtained in Eq.~\eqref{eq:spectrum}. Substituting into
Eq.~\eqref{eq:dyson_2nd} and exchanging the order of integration, we
find
\begin{align}
    \langle e\vert\mathrm{Tr}_{B,r}\{\hat{\rho}^{(2)}(t)\}\vert e\rangle
    &= -2\epsilon^2 \int_{-\infty}^{\infty}\! \frac{d\omega}{2\pi}
    \int_{-\infty}^{\infty}\! \frac{d\omega'}{2\pi}\;
    S_q(\omega)\, S_B(\omega')\nonumber\\
    &\quad\, \times\mathrm{Re}\Big\{\int_0^t dt'\!\int_0^{t'} dt''\;
    e^{i(\omega - \omega')(t'-t'')}\Big\}.
    \label{eq:double_time_integral}
\end{align}
At times $t$ much longer than the correlation time of the integrand
but still short enough for the perturbative expansion to remain valid,
the double time integral sharpens into a delta function via the
asymptotic identity
\begin{align}
    \lim_{t\to\infty}
    2\!\int_0^t \! dt'\!\int_0^{t'}\! dt''
    e^{i(\omega'-\omega)(t'-t'')}
    = 2\pi t\,\delta(\omega'-\omega).
\end{align}
Applying this to Eq.~\eqref{eq:double_time_integral}, the accumulated
ground-state population grows linearly in time,
\begin{align}
   \langle e\vert\mathrm{Tr}_{B,r}\{\hat{\rho}^{(2)}(t)\}\vert e\rangle
  \xrightarrow{\;t\to\infty}
 -\epsilon^2 t\! \int_{-\infty}^{\infty}\! \frac{d\omega}{2\pi}\;
    S_q(\omega)\, S_B(\omega),
\end{align}
from which we identify the qubit decay rate
\begin{align}
    \Gamma_{e\to g}
    =   \int_{-\infty}^{\infty}\! \frac{d\omega}{2\pi}\;
    S_q(\omega)\, S_B(\omega),
    \label{eq:fgr_rate}
\end{align}
recovering Eq.~\eqref{eq:rate} in the main text. The decay rate is thus set by
the spectral overlap between the qubit emission spectrum and the bath
noise density. This
overlap integral is the central object that connects the
number-splitting physics of $S_q(\omega)$ to experimentally measurable $T_1$ degradation.

\begin{figure*}[htbp]
    \centering
    \includegraphics[width=0.95\textwidth]{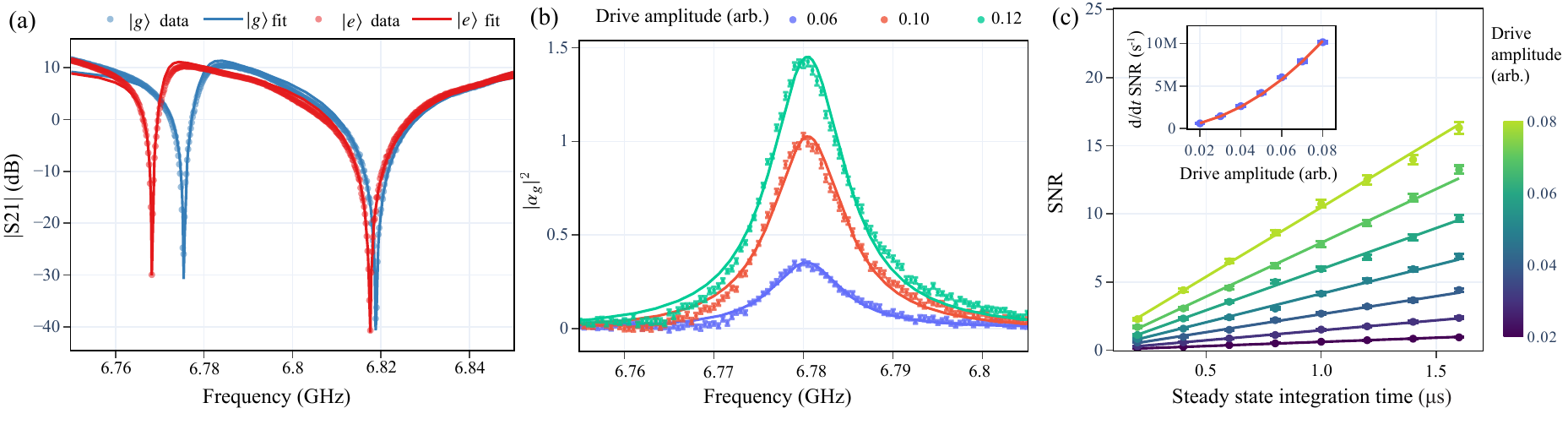}
    \caption{{Readout system characterization.}
    (a) Qubit-state-dependent resonator transmission spectrum $|S_{21}|$ with the transmon prepared in $|g\rangle$ (blue) and $|e\rangle$ (red). Solid lines are fits to a single-mode Purcell-filtered readout model~\cite{Swiadek_2024}. 
    (b) Steady-state intracavity photon number $|\alpha_g|^2$ as a function of drive amplitude at several drive frequencies (colored curves), extracted from the ac-Stark shift of the transmon $g$-$e$ transition measured via qubit absorption spectroscopy.
    (c) Signal-to-noise ratio (SNR) as a function of steady-state integration time for several readout drive amplitudes, demonstrating the linear dependence used to extract the SNR rate, $\frac{d}{dt}\text{SNR}$. Inset: extracted $\frac{d}{dt}\text{SNR}$ as a function of drive amplitude, used to level the measurement rate across different drive detunings. The quantum efficiency of the measurement chain is characterized to be $12.94\,\%$~\cite{Bultink_2018}.}
    \label{fig:readout_characterization}
\end{figure*}

\section{Multi-wave mixing processes}
\label{sec:multi-wave-mixing}
The number splitting of the qubit spectrum can be intuitively interpreted as multi-wave mixing processes. Such processes have been discussed and experimentally observed in other contexts \cite{Astafiev_multi_wave,Zhang_FWM,Gao_FWM,Devoret_DUST,Frattini_twm}, while rarely connected to qubit linewidth broadening under readout. In the following, we establish this connection to explain the emergence of the non-trivial qubit spectrum central to the $T_1$ degradation.

The dressed eigensystem under readout can be derived after performing a so-called polaron transformation \cite{Gambetta_trajectory}
\begin{align}
    \hat{U}_\mathrm{pl} =\exp\Big[\alpha(\hat{n}_q)\hat{a}^\dagger-\mathrm{H.c.}\Big],
  \end{align}
where the qubit-state-dependent resonator displacement is defined by $\alpha(\hat{n}_q)\equiv {d_r}/[\Delta_d-\chi \hat{n}_q+i\kappa/2]$ and we denote $\hat{n}_q\equiv \hat{\sigma}_+\hat{\sigma}_-$ for convenience. For this discussion, we once again neglect the complications introduced by the Purcell filter by setting $\delta_p = 0$ in Eq.~\eqref{eq:Hamiltonian_rot}. After plugging this transform into the Lindblad master equation, we find the following fully diagonalized Hamiltonian \cite{Gambetta_trajectory}
\begin{align}
    \tilde{H} =& \,\omega_q \hat{n}_q + \frac{d_r^2(\Delta_d-\chi\hat{n}_q)}{(\Delta_d-\chi\hat{n}_q)^2+\kappa^2/4} \nonumber\\
    +&\,{\kappa}\mathrm{Im}\{\alpha_e^*\alpha_g\}\hat{n}_q -\Delta_d \hat{a}^\dagger\hat{a} + \chi\hat{n}_q\hat{a}^\dagger\hat{a}
    .\label{eq:polaron_H}
\end{align}
The steady-state resonator displacement $\alpha_e$ and $\alpha_g$ are defined in Eq.~\eqref{eq:alpha_E,G}. In the polaron frame, the pointer states correspond to the displaced vacuum, denoted by $\lvert e, \tilde{0}\rangle$ and $\lvert g, \tilde{0}\rangle$. They are equivalent to the coherent-state pointer states $\lvert e, \alpha_e\rangle$ and $\lvert g, \alpha_g\rangle$ in the rotating frame in which Eq.~\eqref{eq:Hamiltonian_rot} is expressed. 

Eq.~\eqref{eq:polaron_H} gives us a direct way to both compute and understand the AC Stark shift $B$ [c.f.~Eq.~\eqref{eq:corr}] of the qubit: this is just the energy difference of the qubit excited ($n_q = 1$) and ground ($n_q=0$) states with zero photons in the polaron frame.  Recall that $B$ describes the overall shift of the qubit spectrum, and the position of the peak associated with the $n=0$ no-photon transition.  We find  
\begin{align}
    B = &\,\langle e,\tilde{0}\vert \tilde{H}\vert e,\tilde{0}\rangle - \langle g,\tilde{0}\vert \tilde{H}\vert g,\tilde{0}\rangle\nonumber\\
    =&\,\mathrm{Im}\{ i(d_{r}\alpha^*_e-d_{r}\alpha_g) + \kappa \alpha_e^*\alpha_g\}\label{eq:B_again},
\end{align}
which reproduces the shift derived in Eq.~\eqref{eq:B} directly from the number-splitting theory.
The zero-photon energies in the polaron frame include a so-called ``reorganization energy", that is the ground state energy of the conditionally displaced cavity (i.e.~2nd term in Eq.~\eqref{eq:polaron_H}).  More interestingly, there are terms that arise from displacing the cavity loss dissipator (i.e.~third term in Eq.~\eqref{eq:polaron_H} proportional to $\kappa$).  
These different contributions can yield opposing contributions to the overall AC Stark shift $B$.  
Consider the case $\Delta_d = \chi/2$, the standard choice in most experiments for the readout drive frequency (i.e.~$\omega_d = [\omega_r{(e)} + \omega_r{(g)}]/2$).  In this case, the above expression yields an AC Stark shift $B = 4\chi d_r^2(\kappa^2-\chi^2) / (\kappa^2+\chi^2)^2$.  For small $\chi$ this matches the standard perturbative expression $B = \chi \bar{n}$ as expected.  However, for stronger couplings, the dissipative contributions to the Stark shift become subdominant, and for $|\chi|>\kappa$, $B$ changes sign, matching the behavior seen in Fig.~\ref{fig:numerics} (b).  This is a striking departure from weak-coupling expectations.   One can confirm that this behavior only happens for detunings sufficiently close to $\Delta_d = \chi/2$.

To understand the origin of the other peaks, we first check the transformation of the qubit annihilation operator through which the qubit interacts with the environment. Under the polaron transformation, this operator becomes:
\begin{align}
    \hat{\sigma}_- \rightarrow &\,  \hat{U}^\dagger_\mathrm{pl} \hat{\sigma}_-\hat{U}_\mathrm{pl}\nonumber\\
    =&\,\hat{\sigma}_- \exp\Big(\delta\alpha\hat{a}^\dagger - \delta\alpha^*\hat{a}\Big)\nonumber\\
    =&\, \hat{\sigma}_- e^{-|\delta \alpha|^2/2} e^{\delta\alpha \hat{a}^\dagger} e^{-\delta\alpha^* \hat{a}}.\label{eq:splitting}
\end{align}
We now use this transformed operator to describe qubit decay processes.  We assume throughout that the initial state before such decay is stationary: the qubit is in the excited state, and the cavity amplitude has relaxed to its corresponding steady-state amplitude.  In the polaron frame, this means that the initial state is the displaced vacuum $\lvert e, \tilde{0}\rangle$. Therefore, the rightmost operator, $e^{-\delta\alpha^* \hat{a}}$, acts trivially, while the adjacent operator, $e^{\delta\alpha\, \hat{a}^\dagger} = \sum_{j\in\mathbb{N}} (\delta\alpha \hat{a}^\dagger)^n / n!$, maps the vacuum to a number of Fock states in this polaron frame. For nonzero $\delta\alpha$, the qubit–bath interaction can also change the photon number in the resonator. 

The simplest qubit decay process enabled by the dressed operator in \eqref{eq:splitting} involves no photon creation, i.e.~$\vert e,\tilde{0}\rangle \rightarrow \vert g,\tilde{0}\rangle$.  The emission frequency for this process corresponds exactly to the AC Stark shifted qubit frequency, 
c.f.~Eq.~\eqref{eq:B_again}. More non-trivial processes involve qubit decay accompanied by the creation of $n$ photons in the displaced resonator, 
i.e. $\vert e,\tilde{0}\rangle\rightarrow \vert g,\tilde{n}\rangle$, with $n\neq0$ and $\vert g,\tilde{n}\rangle\equiv (\hat{a}^{\dagger})^n \vert g, \tilde{0}\rangle/\sqrt{n!}$. In the displaced frame, the energy cost of adding $n$ photons is \begin{align} \langle g,\tilde{n}\vert \tilde{H}\vert g,\tilde{n}\rangle - \langle g,\tilde{0}\vert \tilde{H}\vert g,\tilde{0}\rangle =-n\Delta_d = n[\omega_r(g)-\omega_d]. \label{eq:gn_energy} \end{align} 
We see that the effective cost of creating a photon is the detuning $\Delta_d$; this reflects the fact that such creation corresponds to conversion of a drive photon to a cavity photon.
We stress crucially that as we are describing qubit decay, the energy of cavity photons in the final state is set by $\omega_r(g)$; there is no dependence on $\omega_r(e)$.

Finally, to make a transition $\vert e,\tilde{0}\rangle \rightarrow \vert g,\tilde{n}\rangle$ resonant, we also need to create a bath excitation with a frequency $\omega_B$ that absorbs the net energy produced in the transition, i.e.~$\omega_B = \tilde{\omega}_q + n\Delta_d$.  This is equivalent to:
\begin{align}
   \tilde{\omega}_q+n\omega_d = \omega_B + n\omega_r(g).
\end{align}
The transition amplitude is determined by $e^{-|\delta\alpha|^2}|\delta\alpha|^2/n!$.

As discussed in the main text, these relatively simple energetic constraints explain why the  special choice of $\omega_d=\omega_r(g)$ minimizes the effective width of the qubit emission spectrum.  For this special choice, the conversion of a drive photon to a final-state cavity photon at $\omega_r(g)$ costs zero energy.  As such, no matter how many cavity photons are produced in a decay event, the generated bath excitation always has an energy $\omega_B = \tilde{\omega}_q $.  
The upshot is that the qubit spectrum for this special detuning $\Delta_d = 0$ always has just a single peak, even for large values of $\chi$, as multi-photon decay processes are not energetically resolved.  



\section{Experimental system parameters and calibration}
\label{app:exp-system}

The device parameters used for the experiment is shown in Table \ref{tab:device_params}. The transmon is flux-tuned to characterize the TLS bath, following which the experiment and calibrations are conducted at the flux bias near the TLS of interest. State preparation, measurement, and reset of the transmon occur at the transmon maximum frequency for the experiment and most calibrations.

To characterize the readout system and dispersive shift, we prepare the transmon in either $|g\rangle$ or $|e\rangle$ then probe the resonator to characterize the qubit-state dependent transmission spectrum. We additionally do a final readout after the probe to post-select against shots where the transmon state transitioned during the resonator probe due to events such as energy relaxation. We fit the resonator transmission spectra to a dedicated single-mode Purcell-filtered model \cite{Swiadek_2024} to extract the relevant readout system parameters [Fig.~\ref{fig:readout_characterization}(a)]. Note that the difference in $|g\rangle$ and $|e\rangle$ resonator linewidths naturally comes from the difference in resonator-Purcell-filter hybridization since the detuning of the two is dependent on the qubit state.

The readout drive strength at the resonator is characterized by ac-Stark shift of the transmon $g$-$e$ transition with variable frequency and amplitude drive on the resonator. We let the resonator come to a steady-state for $>40/\kappa$ before performing qubit-spectroscopy with a relatively short 48 ns $\pi$ pulse -- in this way the complicated number-splitting feature \cite{Gambetta_num_split_theor} of the qubit spectrum reduces to a single Stark-shifted peak in the spectroscopy signal as it is convoluted by a much broader Gaussian envelope. For the scenario where the state-dependent drive strengths are identical $u=v$, the shift is either $|\alpha_e|^2\chi$ for the emission spectrum or $|\alpha_g|^2\chi$ for the absorption spectrum. Unequal drive strengths modify this result slightly. To see this, one can evaluate the first moment of the emission spectrum Eq.~\eqref{eq:spectrum}, i.e., $\int_{-\infty}^{\infty}(d\omega/2\pi)\omega S_q(\omega)$, or similarly for the absorption spectrum as derived in Ref.~\cite{Gambetta_num_split_theor}. This quantity is equivalently expressed as the initial derivative of the correlation function, Eq.~\eqref{eq:full_corr}, giving 
\begin{align}
    \delta\omega_{\mathrm{Stark}} =&\, \mathrm{Im}\{\dot{C}_q(0)\} - \omega_q\nonumber\\
    =&\,|\alpha_e|^2\chi + 2d(u-v)\mathrm{Re}\{\alpha_e\}.\label{eq:stark-shift}
\end{align}
The second term is generally much smaller and vanishes for $\omega_{d}=\omega_r(e)$, where $|\alpha_e|^2$ is maximized. The Stark shift for the absorption spectrum can be derived analogously. The qubit state is finally measured after the resonator rings down. The Stark shift is measured as a function of drive frequency at a few drive amplitudes to calibrate the drive power at the device [Fig.~\ref{fig:readout_characterization}(b)].

We keep the SNR rate leveled at different drive detunings by measuring a calibration curve of $\frac{d}{dt} \text{SNR}$ as a function of drive amplitude over the range of drive detunings [Fig.~\ref{fig:readout_characterization}(c), inset]. The SNR rate is experimentally characterized by measuring the linear dependence of SNR on the integration length when the resonator is in the steady state for a choice of readout drive amplitude and detuning [Fig.~\ref{fig:readout_characterization}(c)]. The SNR for each integration length is obtained from a double Gaussian fit of the readout signal distributions with the transmon prepared in either $|g\rangle$ or $|e\rangle$. Experimentally measured steady state SNR rate can be connected to the measurement rate with the quantum efficiency of the measurement which we characterized to be 12.94 $\%$ with the method described in \cite{Bultink_2018}.

\setlength{\tabcolsep}{6pt} 
\begin{table}
    \centering
    \begin{tabular}{ccc}
        \hline
        \hline
        Transmon frequency&$\omega_q/2\pi$&4746.3 MHz\\
        Dressed resonator frequency&$\omega_r{(g)}/2\pi$&6779.6 MHz\\
        Readout dispersive shift&$\chi/2\pi$&$-8.8$ MHz\\
        Readout linewidth in $|g\rangle$&$\kappa_g/2\pi$&9.0 MHz\\
        Readout linewidth in $|e\rangle$&$\kappa_e/2\pi$&6.6 MHz\\
        Purcell-filter frequency &$\omega_p/2\pi$& 6816.2 MHz\\
        Readout-Purcell coupling & $J/2\pi$& 17.1 MHz\\
        Readout quantum efficiency &$\eta$& 12.94 $\%$\\
        Transmon maximum frequency&$\omega_q^{\text{max}}/2\pi$&5127.3 MHz\\
        Experiment flux bias&$\Phi_{\text{ext}}$ & 0.183 $\Phi_0$\\
        \hline
        \hline
    \end{tabular}
    \caption{Device parameters}
    \label{tab:device_params}
\end{table}


\section{TLS characterization}
\label{app:tls}

\begin{figure}[h]
    \centering
\includegraphics[width=0.65\columnwidth]{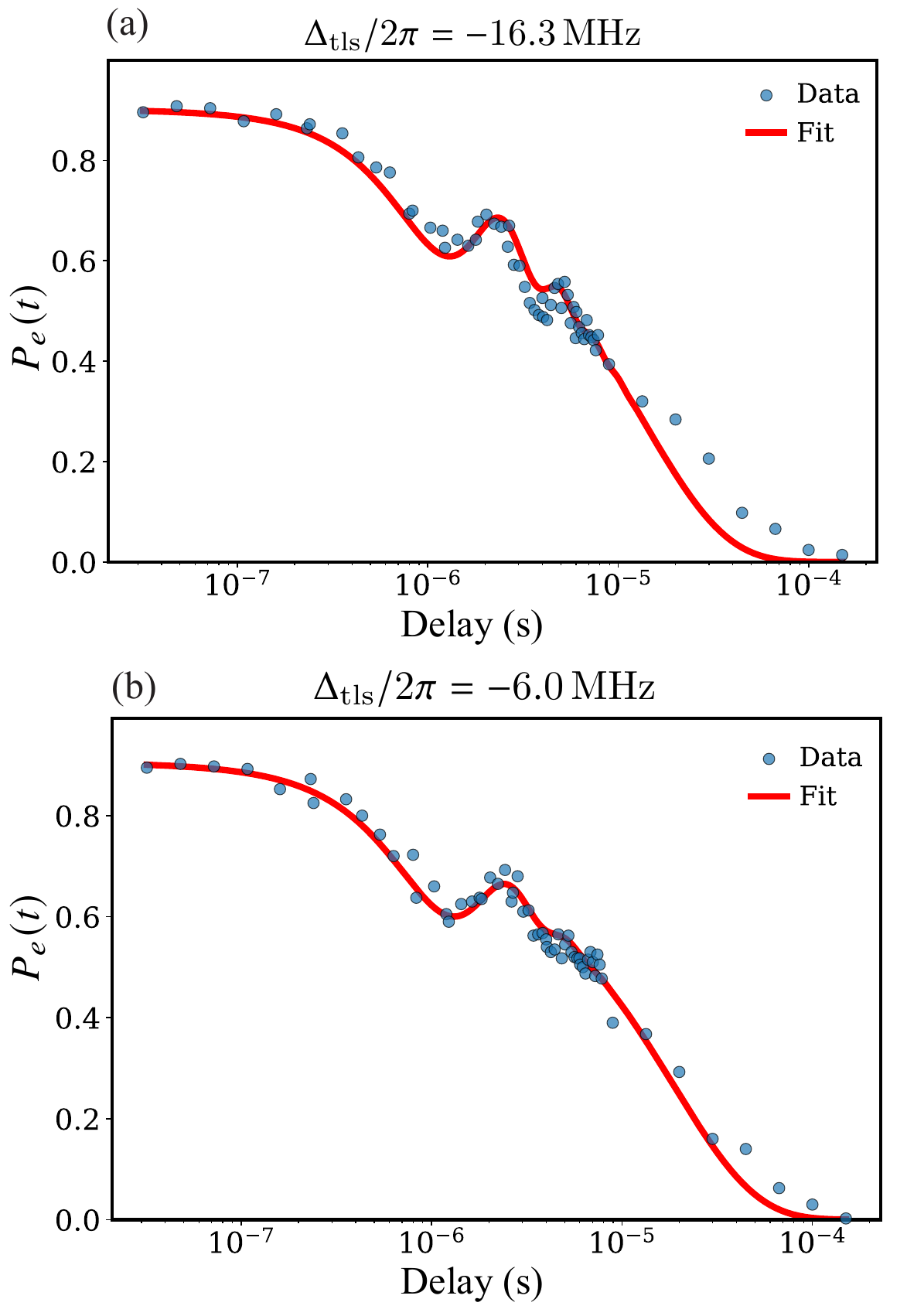}
    \caption{Inversion recovery measurements of the two TLS configurations observed during the experiment. Blue circles: measured excited-state population $P_e(t)$ as a function of delay time; red lines: fits to a double-exponential decay model [Eq.~\eqref{eq:double_exp}]. (a) First TLS configuration: $\Delta_\mathrm{tls}/2\pi = -16.3\,\mathrm{MHz}$, $g_\mathrm{tls}/2\pi = 0.20\,\mathrm{MHz}$, $\gamma_{2,\mathrm{tls}}/2\pi= 0.14\,\mathrm{MHz}$, $\gamma_1/2\pi = 24\,$kHz. (b) Second TLS configuration: $\Delta_\mathrm{tls}/2\pi = -6.0\,\mathrm{MHz}$, $g_\mathrm{tls}/2\pi = 0.19\,\mathrm{MHz}$, $\gamma_{2,\mathrm{tls}}/2\pi= 0.21 \,\mathrm{MHz}$, $\gamma_1/2\pi = 18\,$kHz. The extracted parameters are used to construct the bath spectral density $S_B(\omega)$ entering the theoretical predictions shown in Fig.~\ref{fig:data}.}
    \label{fig:tls_characterization}
\end{figure}

During the experiment, we use the flux tunability of the transmon to 
perform $T_1$ spectroscopy and identify TLS defects in the vicinity of the qubit operating frequency. The qubit idle frequency is chosen to be $366$ MHz below its upper sweet spot. Around that frequency, we identify a single TLS that is both strongly coupled to the transmon and persistent over the 
measurement window, yet switches between two distinct frequency 
configurations on timescales of minutes. We segment the interleaved data by the TLS configuration and analyze each independently.

For a qubit coupled to a single coherent TLS,  the excited-state population can exhibit oscillatory exchange dynamics 
superimposed on exponential decay. We denote coupling strength $g_\mathrm{tls}$, 
detuning $\Delta_\mathrm{tls} \equiv \omega_\mathrm{tls} - \omega_q$, and 
TLS dephasing rate $\gamma_{2,\mathrm{tls}}$. In the regime where the TLS is 
near-resonant with the qubit ($|\Delta_\mathrm{tls}| \ll g_\mathrm{tls}$), the 
coherent swap dynamics between the qubit and TLS produce oscillations at 
frequency $2g_\mathrm{tls}$, damped by the TLS dephasing. The function used to fit the measured decay traces is \cite{muller2009}
\begin{equation}
    P_e(t) = a_1\cos(2g_\mathrm{tls}t)\, e^{-\gamma_{2,\mathrm{tls}} t} 
    + a_2 e^{-\gamma_1 t},
    \label{eq:double_exp}
\end{equation}
where the first term captures the coherent qubit--TLS exchange 
oscillations, and the second term describes the overall exponential 
relaxation of the qubit at rate $\gamma_1$ due to the remaining 
background bath. The coefficients $a_1$ and $a_2$ are treated as free fitting parameters, which further depend on other parameters that are less convenient to measure. 

Fig.~\ref{fig:tls_characterization} shows inversion 
recovery traces for the two TLS configurations, together with fits to 
Eq.~\eqref{eq:double_exp}. The fitted parameters are shown in the caption of the figure. Between the two configurations, the 
coupling strength $g_\mathrm{tls}$ remains approximately unchanged while the TLS decoherence rate $\gamma_{2,\mathrm{tls}}$ and background $\gamma_1$ vary. This is 
consistent with a single defect whose internal dissipation environment 
fluctuates between the two configurations, while its spatial coupling to 
the transmon --- set by the defect's position and dipole moment --- remains fixed.

The extracted parameters are used to construct the bath spectral 
density $S_B(\omega)$ entering Eq.~\eqref{eq:rate}. For 
each configuration, the spectral density is modeled as a Lorentzian \cite{tls_spectrum,you_tls_th}
\begin{align}
    S_{B}(\omega) = 
    \frac{2g_{\mathrm{tls}}^2 \gamma_{2,\mathrm{tls}}}
    {(\omega - \omega_\mathrm{tls})^2 + \gamma_{2,\mathrm{tls}}^2} + 1/T_1,
    \label{eq:tls_spectrum}
\end{align}
with the offset $1/T_1$ accounting for offset qubit decay rate contributed by other environmental interactions other than this prominent TLS. The form of Eq.~(\ref{eq:tls_spectrum}) assumes the TLS resides in its ground state in the absence of qubit-induced excitation. This is justified here because the readout drive is applied to the resonator and is far detuned from the TLS transitions, so the drive is both spatially decoupled and spectrally detuned from the TLS. Furthermore, we allow sufficient relaxation of the system to thermal equilibrium between shots. Thermal excitation of the TLS is also negligible: at the TLS frequencies relevant here ($\omega_\mathrm{tls}/2\pi \approx 5 $ GHz) and typical dilution-refrigerator temperatures ($\approx20$~mK), the corresponding thermal occupation is exponentially suppressed, and finite-temperature corrections to $S_B(\omega)$ can therefore be safely neglected.

This procedure allows us to accurately construct the bath spectrum entering the qubit decay rate in Eq.~\eqref{eq:rate}, enabling a direct comparison between the measured readout-induced $T_1$ degradation and the theoretical predictions shown in Fig.~\ref{fig:data}.

\bibliography{mybib}

\end{document}